\documentclass[preprint,journal]{vgtc}       

\ifpdf 
    \pdfoutput=1\relax 
    \usepackage{graphicx} 
    \DeclareGraphicsExtensions{.pdf,.png,.jpg,.jpeg} 
\else 
    \usepackage{graphicx} 
    \DeclareGraphicsExtensions{.eps} 
\fi

\graphicspath{{figures/}{images/}{./}} 

\usepackage{microtype} 
\PassOptionsToPackage{warn}{textcomp} 
\usepackage{textcomp} 

\usepackage{mathptmx} 
\usepackage{times} 
\usepackage{cite} 
\usepackage{tabu} 
\usepackage{booktabs} 

\vgtcinsertpkg

\usepackage{stfloats}
\usepackage{titlesec}
\usepackage{amsmath}
\usepackage{gensymb}
\usepackage{amssymb}
\usepackage{enumitem}
\usepackage[usenames, dvipsnames]{color}
\usepackage[dvipsnames]{xcolor}
\usepackage[caption=false]{subfig}
\usepackage{calrsfs}
\usepackage{tabularx}
\usepackage{float}
\usepackage{wrapfig}
\usepackage{setspace}
\usepackage[keeplastbox]{flushend}

\usepackage[T1]{fontenc}
\usepackage[scaled=0.825]{beramono}
\usepackage{anyfontsize}

\DeclareMathAlphabet{\pazocal}{OMS}{zplm}{m}{n}

\newcommand\RemovePeriod[1]{}

\newcommand{\textsfsm}[1]{{\fontsize{8.25}{9.9}\textsf{#1}}}
\newcommand{\textsfverysm}[1]{{\fontsize{7.5}{8}\textsf{#1}}}
  
\preprinttext{\parbox{0.9\textwidth}{\vspace{-5pt} \centering {\fontsize{6.5}{7} \selectfont $\copyright$ \the\year~IEEE. This is the author's version of the article that has been published in IEEE Transactions on Visualization and Computer Graphics.\\ \vspace{-2pt} The published version of this article is available at: \href{https://doi.org/10.1109/TVCG.2019.2934396}{10.1109/TVCG.2019.2934396}}}}

\onlineid{1008}

\vgtccategory{Research}
\vgtcpapertype{algorithm/technique}

\title{A Deep Generative Model for Graph Layout\ifpreprint\else\vspace{-0.2em}\fi}

\author{Oh-Hyun~Kwon and Kwan-Liu~Ma}
\authorfooter{
\ifpreprint\else\vspace{-0.85em}\fi
\item The authors are with the University of California, Davis.\protect\\E-mail: kw@ucdavis.edu, ma@cs.ucdavis.edu.
}

\shortauthortitle{Kwon and Ma: A Deep Generative Model for Graph Layout}

\abstract{Different layouts can characterize different aspects of the same graph.
Finding a ``good'' layout of a graph is thus an important task for graph visualization. 
In practice, users often visualize a graph in multiple layouts by using different methods and varying parameter settings until they find a layout that best suits the purpose of the visualization.
However, this trial-and-error process is often haphazard and time-consuming.
To provide users with an intuitive way to navigate the layout design space, we present a technique to systematically visualize a graph in diverse layouts using deep generative models.
We design an encoder-decoder architecture to learn a model from a collection of example layouts, where the encoder represents training examples in a latent space and the decoder produces layouts from the latent space.
In particular, we train the model to construct a two-dimensional latent space for users to easily explore and generate various layouts.
We demonstrate our approach through quantitative and qualitative evaluations of the generated layouts.
The results of our evaluations show that our model is capable of learning and generalizing abstract concepts of graph layouts, not just memorizing the training examples.
In summary, this paper presents a fundamentally new approach to graph visualization where a machine learning model learns to visualize a graph from examples without manually-defined heuristics.
}


\teaser{
\ifpreprint\vspace{-6pt}\else\vspace{-8pt}\fi
\centering
\captionsetup{farskip=0pt,skip=0pt}
\includegraphics[width=\textwidth,keepaspectratio]{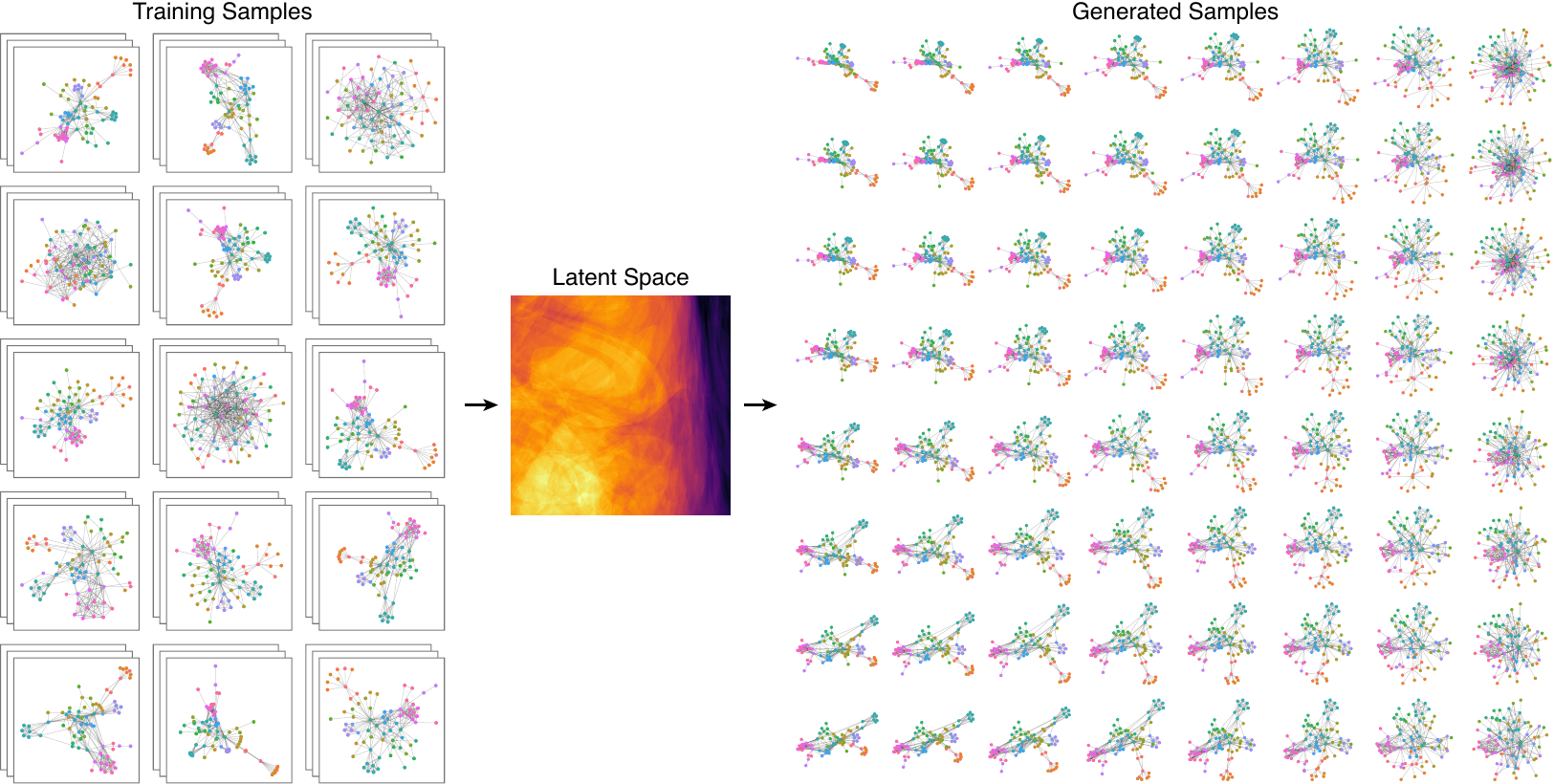}
\vspace{-11pt}
\caption{Generative modeling of layouts for the \textit{Les Mis\'{e}rables} character co-occurrence network \cite{StandfordGraphBase}.
We train our generative model to construct a 2D latent space by learning from a collection of example layouts (training samples).
From the grid of generated samples, we can see the smooth transitions between the different layouts.
This shows that our model is capable of learning and generalizing abstract concepts of graph layouts, not just memorizing the training samples.
Users can use this sample grid of the latent space as a WYSIWYG interface to generate a layout they want, 
without either blindly tweaking parameters of layout methods or requiring expert knowledge of layout methods.
The color mapping of the latent space represents the shape-based metric \cite{Eades17} of the generated samples.
Throughout the paper, unless otherwise specified, the node color represents the hierarchical community structure of the graph \cite{Peixoto14, TreeColors}, so readers can easily compare node locations in different layouts.
An interactive demo is available in the supplementary material \cite{Supplementary}.}
\label{fig:teaser}
\ifpreprint\else\vspace{-3pt}\fi
}

\nocopyrightspace

\begin{document}


\firstsection{Introduction}
\maketitle
Graphs are commonly used for representing complex systems, such as interactions between proteins, data communications between computers, and relationships between people.
Visualizing a graph can help better understand the relational and structural information in the data that would not be as apparent if presented in a numeric form.
The most popular and intuitive way to visualize a graph is a node-link diagram, where the nodes are drawn as points, and the links are rendered as lines.

Drawing a node-link diagram by hand is laborious; since the 1960s, researchers have devised a multitude of methods to automatically lay out a graph.
The layouts of the same graph can vary greatly depending on which method is used and the method's configuration.
However, there is no ``best'' layout of a graph as different layouts often highlight different structural characteristics of the same graph \cite{Blythe96, Dunne09}.
For example, while one layout can emphasize connections between different communities of a graph, it might not be able to depict connections within each community.
Thus, it is important to find a ``good'' layout for showing the features of a graph that users want to highlight.

Finding a good layout of a graph is, however, a challenging task.
The heuristics to find a good layout are nearly impossible to define. 
It requires to consider many different graphs, characteristics to be highlighted, and user preferences. 
There is thus no existing method to automatically find a good layout.
In practice, users rely on a trial-and-error process to find a good layout.
Until they find a layout that satisfies their requirements (e.g., highlighting the community structure of a graph), users typically visualize a graph in multiple layouts using different methods and varying parameter settings.
This process often requires a significant amount of the user's time as it results in a haphazard and tedious exploration of a large number of layouts \cite{Biedl98}.

Furthermore, expert knowledge of layout methods is often required to find a good layout.
Most layout methods have a number of parameters that can be tweaked to improve the layout of a graph.
However, many layout methods---especially force-directed ones---are very sensitive to the parameter values \cite{Munzner14:VAD}, where the resulting layouts can be incomprehensible or even misleading \cite{Dunne09}.
A proper selection of the parameter settings for a given graph requires detailed knowledge of the chosen method.
Such knowledge can only be acquired through extensive experience in graph visualization.
Thus, novice users are often blindly tweaking parameters, which leads to many trials and errors as they cannot foresee what the resulting layout will look like.
Moreover, novices might explore only a fraction of possible layouts, choose an inadequate layout, and thus overlook critical insights in the graph.

To help users to produce a layout that best suits their requirements, we present a deep generative model that systematically visualizes a graph in diverse layouts.
We design an encoder-decoder architecture to learn a generative model from a collection of example layouts, where the encoder represents training examples in a latent space and the decoder generates layouts from the latent space.
In particular, we train the model to construct a two-dimensional latent space.
By mapping a grid of generated samples, a two-dimensional latent space can be used as a what-you-see-is-what-you-get (WYSIWYG) interface. 
This allows users to intuitively navigate and generate various layouts without blindly tweaking parameters of layout methods.
Thus, users can create a layout that satisfies their requirements without a haphazard trial-and-error process or any expert knowledge of layout methods.

The results of our evaluations show that our model is capable of learning and generalizing abstract concepts of graph layouts, not just memorizing the training examples.
Also, graph neural networks \cite{Xu19GIN, Kipf16GCN} and Gromov--Wasserstein distance \cite{Memoli11} help the model better learn the complex relationship between the structure and the layouts of a graph.
After training, our model generates new layouts considerably faster than existing layout methods. 
In addition, the generated layouts are spatially stable, which helps users to compare various layouts.

In summary, we introduce a fundamentally new approach to graph visualization, where a machine learning model learns to visualize a graph as a node-link diagram from existing examples without manually-defined heuristics.
Our work is an example of \textit{artificial intelligence augmentation} \cite{AIA}: a machine learning model builds a new type of user interface (i.e., layout latent space) to augment a human intelligence task (i.e., graph visualization design).
\section{Related Work}
Our work is related to graph visualization, deep generative modeling, and deep learning on graphs.
We discuss related work in this section.

\subsection{Graph Visualization}
\label{sec:rel:graphvis}
A plethora of layout methods has been introduced over the last five decades.
In this paper, we focus on two-dimensional layout methods that produce straight-edge drawings, such as force-directed methods \cite{Eades84, Fruchterman91, Frick94, Kamada89, Davidson96}, 
dimensionality reduction-based methods \cite{Brandes06, Harel04, Kruiger17},
spectral methods \cite{Civril05, Koren05},
and multi-level methods \cite{Walshaw03, Gajer02, Harel02, FM3, Hu05, Frishman07}.
These layout methods can be used for visualizing any type of graphs.
Many analysts do not have expert knowledge in these layout methods for finding a good layout.
Our goal is to learn the layouts produced by the different methods in a single model.
In addition, we provide an intuitive way for users to explore and generate diverse layouts of a graph without the need of expert knowledge of layout methods.

How to effectively find a ``good'' layout of a graph is still an open problem.
However, decades of research in this area have led to several heuristics, often called \emph{aesthetic criteria}, for improving and evaluating the quality of a layout.
For instance, \emph{reducing edge crossings} has been shown as one of the most effective criteria to improve the quality of a layout \cite{Kieffer16, Purchase12, Huang07, Purchase02C}.
Based on the aesthetic criteria, several metrics have been defined to evaluate layouts quantitatively.

However, even with aesthetic criteria and metrics, human intervention is still needed in the process of selecting a layout for several reasons.
First, while each heuristic attempts to enhance certain aspects of a layout,
it does not guarantee an overall quality improvement \cite{Dunne09, Gibson12}.
For instance, depending on the given circumstances (e.g., given graph, task, and environment),
certain criteria can lead to incomprehensible layouts \cite{Blythe96}.
Second, there is no consensus on which criteria are the most helpful in a given circumstance \cite{Dunne09, Gibson12, Kieffer16}.
Third, it is often not feasible to satisfy several aesthetic criteria in one layout because satisfying one may result in violating others \cite{DiBattista94, DiBattista98}.
Lastly, selecting a good layout is highly subjective, as each person might have varying opinions on what is a ``good'' layout.
For these reasons, users often rely on a trial-and-error process to find a good layout.

To help users find a good layout, several methods have been developed to accelerate the trial-and-error selection process by learning user preferences \cite{Masui94, Biedl98, Barbosa01, Sponemann14}.
For example, Biedl et al. \cite{Biedl98} have introduced the concept of \textit{multidrawing}, which systematically produces many different layouts of the same graph.
Some methods use an evolutionary algorithm \cite{Masui94, Barbosa01, Sponemann14} to optimize a layout based on a human-in-the-loop assessment.
However, these methods require constant human intervention throughout the optimization process.
In addition, the goal of their optimization is to narrow down the search space. This allows the model to create only a limited number of layouts.
Hence, multiple learning sessions might be needed to allow users to investigate other possible layouts of the same graph.
In contrast to these models, our approach is to train a machine learning model in a fully unsupervised manner to produce diverse layouts, not to narrow the search space.

Recently, several machine learning approaches have been introduced to different tasks in graph visualization, such as previewing large graphs \cite{kwon18wgl}, exploring large graphs~\cite{Chen19StructExplore}, and evaluating visualizations~\cite{Klammler18, Haleem18}.
Unlike these approaches, our goal is to train a model to generate layouts.

\subsection{Deep Generative Models}
The term ``generative model'' can be used in different ways.
In this paper, we refer to a model that can be trained on unlabeled data and is capable of generating new samples that are similar to, but not the same as, the training data.
For example, we can train a model to create synthetic images of handwritten digits by learning from a large collection of real ones \cite{VAE, GAN}.
Since generating new, realistic samples requires a good understanding of the given data,
generative modeling is often considered as a key component of unsupervised learning.

In recent years, generative models built with deep neural networks and stochastic optimization methods have demonstrated state-of-the-art performance in various applications, such as
text generation~\cite{Hu17TextGeneration},
music generation~\cite{MusicVAE},
and drug design~\cite{GomezBombarelli18}.
While several approaches have been proposed for deep generative modeling,
the two most prominent ones are variational autoencoders (VAEs)~\cite{VAE} and generative adversarial networks (GANs)~\cite{GAN}.

VAEs and GANs have their own advantages and disadvantages.
GANs generally produce visually sharper results when applied to an image dataset as they can implicitly model a complex distribution.
However, training GANs is difficult due to non-convergence and mode collapse~\cite{GANTutorial}.
VAEs are easier to train and provide both a generative model and an inference model.
However, they tend to produce blurry results when applied to images.

For designing our generative model, we use sliced-Wasserstein autoencoders (SWAEs) \cite{SWAE}.
As a variant of VAE, it is easier to train than GANs.
In addition, it is capable of learning complex distributions. 
SWAEs allow us to shape the distribution of the latent space into any samplable probability distribution without training an adversarial network or defining a likelihood function.

\subsection{Deep Learning on Graphs}
Machine learning approaches to graph-structured data, such as social networks and biological networks, require an effective representation of the graph structure.
Recently, many graph neural networks (GNNs) have been proposed for representation learning on graphs, such as graph convolutional networks~\cite{Kipf16GCN}, GraphSAGE~\cite{GraphSAGE}, and graph isomorphism networks~\cite{Xu19GIN}.
They have achieved state-of-the-art performance for many tasks,  such as graph classification, node classification, and link prediction.
We also use GNNs for learning the complex relationship between the structure and the layouts of a graph.

Several generative models have been introduced for graph-structured data using GNNs~\cite{MolGAN, GraphRNN, GraphVAE, Graphite, Ma18RVAE}.
These models learn to generate whole graphs for tasks that require new samples of graphs, such as \emph{de novo} drug design.
However, generating a graph layout is a different type of task, where the structure of a graph remains the same, but the node attributes (i.e., positions) are different.
Thus, we need a model that learns to generate different node attributes of the same graph.
\section{Approach}
\label{sec:approach}
Our goal is to learn a generative model that can systematically visualize a graph in diverse layouts from a collection of example layouts.
We describe the entire process for building a deep generative model for graph layout, from collecting training data to designing our architecture.

\subsection{Training Data Collection}
\label{sec:approach:training_data}
Learning a generative model using deep neural networks requires a large amount of training data.
As a data-driven approach, the quality of the training dataset is crucial to build an effective model.
For our goal, we need a large and diverse collection of layouts of the input graph. 

Grid search is often used for parameter optimization \cite{RandomParamSearch}, where a set of values is selected for each parameter, and the model is evaluated for each combination of the selected sets of values.
It is often used for producing multiple layouts of a graph. 
For example, Haleem et al. \cite{Haleem18} have used grid search for producing their training dataset.
However, the number of combinations of parameters increases exponentially with each additional parameter.
In addition, different sets of parameter values are often required for different graphs.
Therefore, it requires expert knowledge of each layout method to carefully define the search space for collecting the training data in a reasonable amount of time.

We collect training example layouts using multiple layout methods following random search, where each layout is computed using randomly assigned parameter values of a method.
We uniformly sample a value from a finite interval for a numerical parameter or a set of possible values for a categorical parameter.
Random search often outperforms grid search for parameter optimization \cite{RandomParamSearch}, especially when only few parameters affect the final result.
Because the effect of the same parameter value can vary greatly depending on the structure of the graph, we believe random search would produce a more diverse set of layouts than grid search.
Moreover, for this approach, we only need to define the interval of values for a numeric parameter, which is a simpler task for non-experts than selecting specific values for grid search.

Computing a large number of layouts would take a considerable amount of time.
However, we can start training a model without having the full training dataset, thanks to stochastic optimization methods.
For example, we can train a model with stochastic gradient descent \cite{SGD}, where a small batch of training examples (typically a few dozens) are fed to the model at each step.
Thus, we can incrementally train the model while generating the training examples simultaneously.
This allows the users to use our model as early as possible.

\subsection{Layout Features}
\label{sec:layout-feature}
Selecting informative, discriminating, and independent features of the input data is also an essential step for building an effective machine learning model.
With deep neural networks, we can use low-level features of the input data without handcrafted feature engineering.
For example, the red, blue, and green channel values of pixels often are directly used as the input feature of an image in deep learning models.

However, what would be a good feature of a graph layout?
Although the node positions can be a low-level feature of a layout, using the raw positions as a feature has several issues.

Many graph layout methods do not use spatial position to directly encode attribute values of either nodes or edges.
The methods are designed to optimize a layout following certain heuristics, such as 
minimizing the difference between the Euclidean and graph-theoretic distances, reducing edge crossings, and minimizing node overlaps.
Therefore, the position of a node is often a side effect of the layout method; it does not directly encode any attributes or structural properties of a node \cite{Munzner14:VAD}.

In addition, many layout methods are nondeterministic since they employ randomness in the layout process to avoid local minima, such as randomly initializing the positions of nodes \cite{Munzner14:VAD}.
Therefore, the position of a node can significantly vary between different runs of the same layout method with the same parameter setting.
For these reasons, the node positions are not a reliable feature of a layout.

Besides, due to the Gestalt principle of proximity \cite{Wertheimer1923:Gestalt}, the nodes placed close to each other would be perceived as a group whether or not this relationship exists \cite{McGrath96}.
For example, some nodes might be placed near each other because they are pushed away from elsewhere, not because they are closely connected in the graph \cite{Munzner14:VAD}.
However, the viewers can perceive them as a cluster because spatial proximity strongly influences how the viewer perceives the relationships in the graph \cite{Gibson12}.
Thus, spatial proximity is an essential feature of a layout.

We use the pairwise Euclidean distance of nodes in a layout as the feature of the layout.
As the spatial proximity of nodes is an intrinsic feature of a layout,
we can directly use the pairwise distance matrices to compare different layouts, without considering the rotation or reflection of the points.
Furthermore, we can normalize a pairwise distance matrix of nodes by its mean value for comparing layouts in different scales.

Therefore, for the positions of the nodes ($P$) of a given layout, we compute the feature of the given layout by $X_L = D / \bar{D}$, where $D$ is the pairwise distance matrix of $P$ and $\bar{D}$ is the mean of $D$.
Each row of $X_L$ is the node-level feature of a layout for each node.
Other variations of the pairwise distance can be used as a feature, such as the Gaussian kernel from the pairwise distance: $\exp\left(-D^2/2\sigma\right)$.
\vspace{-2pt}

\subsection{Structural Equivalence}
\label{sec:struct_eq}
Two nodes of a graph are said to be structurally equivalent if they have the same set of neighbors.
Structurally equivalent nodes (SENs) of a graph are often placed at different locations, as many layout methods have a procedure to prevent overlapping.
For example, force-directed layout methods apply repulsive forces between all the nodes of a graph.
As the layout results are often nondeterministic, 
the positions of SENs are mainly determined by the randomness in the layout method.

\begin{wrapfigure}[9]{r}{50pt}
\vspace{-20pt}
\begin{center}
\includegraphics[]{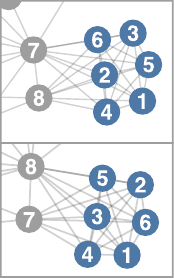}
\end{center}
\end{wrapfigure}
\noindent
For instance, the figure on the right shows two different layout results of the same graph using the same layout method (sfdp \cite{Hu05}) with the same default parameter setting. 
The blue nodes are the same set of SENs.
However, the arrangements of the blue nodes between the two layouts are quite different. 
In the top layout, the nodes $\{6, 2, 4\}$ are closer to $\{7, 8\}$ than $\{3, 5, 1\}$. In contrast, in the bottom layout, the nodes $\{5, 3, 4\}$ are closer to $\{7, 8\}$ than $\{2, 6, 1\}$.
This presents a challenge because a permutation invariant measure of similarity is needed for the same set of SENs.
In other words, we need a method to permute the same set of SENs, from one possible arrangement to the other, for a visually correct similarity measure.

To address this issue, we compare two different layouts of the same set of SENs using the Gromov--Wasserstein (GW) distance \cite{Memoli11, Peyre16}:
\vspace{-4.5pt}
\begin{equation}
\operatorname{GW}(C, C') = \min_{T} \sum_{i,j,k,l} L\left(C_{i,k},C'_{j,l}\right)\ T_{i,j}\ T_{k,l},
\vspace{-5.5pt}
\label{eq:gw}
\end{equation}
where $C$ and $C'$ are cost matrices representing either similarities or distances between the objects of each metric space, 
$L$ is a loss function that measures the discrepancy between the two cost matrices (e.g., $L_2$ loss),
and $T$ is a permutation matrix, which couples the two metric spaces, that minimizes $L$.
The GW distance measures the difference between two metric spaces.
For example, it can measure the dissimilarity between two point clouds, invariant to the permutations of the points.

For a more efficient computation, 
we do not backpropagate through the GW distance in the optimization process of our model.
Instead, we use the permutation matrix $T$ to permute the same set of SENs.
We describe this in more detail in \autoref{sec:approach:training}

\begin{figure*}[t]
\captionsetup{farskip=0pt,skip=0pt,captionskip=0pt}
\centering
\includegraphics[]{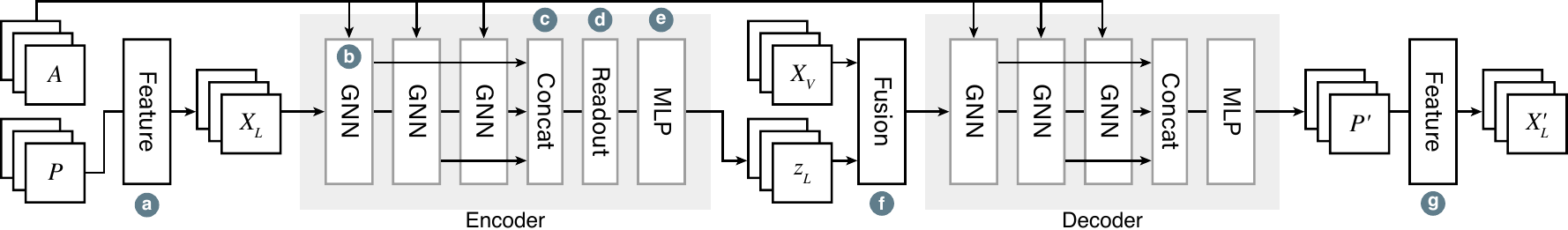}
\vspace{-10pt}
\caption{%
Our encoder-decoder architecture that learns a generative model from a collection of example layouts.
We describe it in \autoref{sec:approach}}
\label{fig:architecture}
\vspace{-12pt}
\end{figure*}

\subsection{Architecture}
\label{sec:arch}
We design an encoder-decoder architecture that learns a generative model for graph layout.
Our architecture and optimization process generally follow the framework of VAEs \cite{VAE}.
The overview of our architecture is shown in \autoref{fig:architecture}.

\paragraph{\textbf{Preliminaries}}
An autoencoder (AE) learns to encode a high-dimensional input object to a lower-dimensional latent space and then decodes the latent representation to reconstruct the input object.
A classical AE is typically trained to minimize \textit{reconstruction loss}, which measures the dissimilarity between the input object and the reconstructed input object.
By training an AE to learn significantly lower-dimensional representations than the original dimensionality of the input objects, the model is encouraged to produce highly compressed representations that capture the essence of the input objects.

A classical AE does not have any regularization of the latent representation. 
This leads to an arbitrary distribution of the latent space, which makes it difficult to understand the shape of the latent space.
Therefore, if we decode some area of the latent space, we would get reconstructed objects that do not look like any of the input objects, as that area has not been trained for reconstructing any input objects.

VAEs \cite{VAE} extend the classical AEs by minimizing \textit{variational loss}, which measures the difference of the distribution of the input objects' latent representations and a \textit{prior} distribution.
Minimizing the variational loss encourages VAEs to learn the latent space that follows a predefined structure (i.e., prior distribution).
This allows us to know which part of the latent space is trained with some input objects.
Thus, it is easy to generate a new object similar to some of the input objects.

\paragraph{\textbf{Encoder}}
For our problem, the input objects are graph layouts, i.e., the positions of nodes ($P$, each row is the position of a node).
We first compute the feature of a layout ($X_L$) 
as discussed in \autoref{sec:layout-feature} (\autoref{fig:architecture}a), where each row corresponds to the input feature of a node.

The encoder takes the feature of a layout ($X_L$) and the structure of a graph ($A$, the adjacency matrix), as shown in \autoref{fig:architecture}b. Then, it outputs the latent representation of a layout ($z_L$).
We use graph neural networks (GNNs) to take the graph structure into account in the learning process.

In general, GNNs learn the representation of a node following a recursive neighborhood aggregation (or message passing) scheme, 
where the representation of a node is derived by recursively aggregating and transforming the representations of its neighbors.
For instance, graph isomorphism networks (GINs) \cite{Xu19GIN} update node representations as:
\vspace{-3pt}
\begin{equation}
h_v^{(k)}\!= \operatorname{MLP}^{(k)}\!\left(\!( 1 + \epsilon^{(k)} )\!\cdot h_v^{(k-1)}\!+ 
f^{(k)}\!\left(\!\left\{ h_u^{(k-1)}\!:\!u \in \pazocal{N}(v) \right\}\!\right)
\right),
\vspace{-6pt}
\label{eq:gin}
\end{equation}
where 
$h_v^{(k)}$ is the feature vector of node $v$ at the $k$-th aggregation step
($h_v^{(0)}$ is the input node feature),
$\pazocal{N}(v)$ is a set of neighbors of node $v$,
$f$ is a function that aggregates the representations of $\pazocal{N}(v)$, such as element-wise mean,
$\epsilon$ is a learnable parameter or a fixed scalar that weights the representation of node $v$ and the aggregated representation of $\pazocal{N}(v)$, and
MLP is a multi-layer perceptron to learn a function that transforms node representations.
After $k$ steps of aggregation, our encoder produces the latent representation of a node, which captures the information within the node's $k$-hop neighborhood. 

The aggregation scheme of GNNs corresponds to the subtree structure rooted at each node, where it learns a more global representation of a graph as the number of aggregations steps increases \cite{Xu19GIN}.
Therefore, depending on the graph structure, an earlier step may learn a better representation of a node.
To consider all representations at different aggregation steps, we use the outputs of all GNN layers.
This is achieved by concatenating the outputs of all GNN layers (\autoref{fig:architecture}c) similar to \cite{JumpKnow}.

The graph-level representation of a layout is obtained with a readout function (\autoref{fig:architecture}d).
A readout function should be permutation invariant to learn the same graph-level representation of a graph, regardless of the ordering of the nodes.
It can be a simple element-wise mean pooling or a more advanced graph-level pooling, such as DiffPool \cite{DiffPool}.
We use MLP to produce the final output representation of a layout $z_L$ (\autoref{fig:architecture}e).

Although we could use a higher-dimensional space, we construct a 2D latent space since users can intuitively navigate the latent space.
By mapping a grid of generated samples, a 2D latent space can be used as a WYSIWYG interface (more in \autoref{sec:scenario}).
For this, we set the prior distribution as the uniform distribution in $[-1, 1]^2$.
Thus, the encoder produces a 2D vector representation of a layout ($z_L$) in $[-1, 1]^2$.

\paragraph{\textbf{Fusion Layer}}
The encoder produces a graph-level representation of each layout ($z_L$).
If we only use this graph-level representation, all nodes will have the same feature value for the decoder.
To distinguish the individual nodes, we use one-hot encoding of the nodes (i.e., identity matrix) as the feature of the nodes ($X_V$), similar to the featureless case of GCN \cite{Kipf16GCN}.
Then, we combine the graph-level representation of a layout ($z_L$) and node-level features ($X_V$) using a fusion layer \cite{Iizuka16} (\autoref{fig:architecture}f).
It fuses $z_L$ and $X_V$ by concatenating each row of $X_V$ with $z_L$.

\paragraph{\textbf{Decoder}}
The decoder takes the fused features and learns to reconstruct the input layouts, i.e., it reconstructs the position of the nodes ($P'$).
The decoder has a similar architecture to the encoder, except that it does not have a readout function, as the output of the decoder is a node-level representation, i.e., the positions of the nodes ($P'$).
The feature of the reconstructed layout ($X'_L$) is computed for measuring the reconstruction loss (\autoref{fig:architecture}g).
After training, users can generate diverse layouts by feeding different $z_L$ values to the decoder.

\subsection{Training}
\label{sec:approach:training}
Following the framework of VAEs \cite{VAE}, 
we learn the parameters of the neural network used in our model by minimizing the reconstruction loss ($L_X$) and the variational loss ($L_Z$).

The reconstruction loss ($L_X$) measures the difference between the input layout ($P$) and its reconstructed layout ($P'$).
As we have discussed in \autoref{sec:layout-feature}, we compare the features of the two layouts ($X$ and $X'$) to measure the dissimilarity between them.
For example, we can use the $L_1$ loss function between the two layouts $L_X = \|X_L - X'_L\|_1$.

As we have discussed in \autoref{sec:struct_eq}, for the same set of structurally equivalent nodes (SENs) in a graph, we use the Gromov--Wasserstein (GW) distance between the two layouts ($P$ and $P'$) for the comparison.
However, for a more efficient computation, we do not backpropagate through the GW distance in the optimization process.
Computing GW yields a permutation matrix ($T$ in \autoref{eq:gw}), 
as it is based on ideas from mass transportation \cite{Villani03}.
We use this permutation matrix to permute the input layout $\hat{P} = T P$ and compute the feature of the permuted input layout $\hat{X}_L$.
Then, we compute the reconstruction loss between the permuted input layout ($\hat{P}$) and the reconstructed layout ($P'$). 
For example, if we use the $L_1$ loss, we can compute the reconstruction loss as $L_X = \|\hat{X_L} - X'_L\|_1$.
With this method, we can save computation cost for the backpropagation through the complex GW computation, and we still can compare the different layouts of the SENs properly, without affecting the model performance.

In this work, we use the variational loss function defined in sliced-Wasserstein autoencoders (SWAE) \cite{SWAE}:
\vspace{-5pt}
\begin{equation}
\operatorname{SW}_{\!c}\!\left(p, q \right) = \frac{1}{L\!\cdot\!M} \sum_{l=1}^{L} \sum_{m=1}^{M} c\!\left(\theta_l \cdot p_{i\left[m\right]},\ \theta_l \cdot q_{j\left[m\right]} \right),
\vspace{-5pt}
\label{eq:sw}
\end{equation}
where 
$p$ and $q$ are the samples from two distributions,
$M$ is the number of the samples, 
$\theta_l$ are random \textit{slices} sampled from a uniform distribution on a $d$-dimensional unit sphere ($d$ is the dimension of $p$ and $q$),
$L$ is the number of random slices, 
$i\left[m\right]$ and $j\left[m\right]$ are the indices of sorted $\theta_l \cdot p_{i\left[m\right]}$ and $\theta_l \cdot q_{j\left[m\right]}$ with respect to $m$, correspondingly, $c$ is a transportation cost function (e.g., $L_2$ loss).

Using sliced-Wasserstein distance for variational loss allows us to shape the distribution of the latent space into any samplable probability distribution without defining a likelihood function or training an adversarial network.
The variational loss of our model is $L_Z = \operatorname{SW}_c(Z_L, Z_P)$, where $Z_P$ is a set of samples of the prior distribution.

The optimization objective can be written as:
\vspace{-5pt}
\begin{equation}
\operatorname*{argmin}_{E_{\theta}, D_{\phi}}~L_X + \beta L_Z,
\vspace{-8pt}
\label{eq:loss}
\end{equation}
where $E_{\theta}$ is the encoder parameterized by $\theta$, $D_{\phi}$ is the decoder parameterized by $\phi$, and $\beta$ is the relative importance of the two losses.
\section{Evaluation}
\label{sec:eval}
The main goal of our evaluations is to see whether our model is capable of learning and generalizing abstract concepts of graph layouts, not just memorizing the training examples.
We perform quantitative and qualitative evaluations of the reconstruction of unseen layouts (i.e., the test set), and a qualitative evaluation of the learned latent space.

\begin{table}[t]
\setlength{\tabcolsep}{0.175em}
\captionsetup{farskip=0pt,skip=3pt}
\centering
\caption{The nine graphs used in the evaluation. 
$|V|$:~the number of nodes, 
$|E|$:~the number of edges, 
$|S|$:~the number of nodes having structural equivalence,
$l$:~average path length.}
{\scriptsize
\begin{tabu}{@{}l l r r r r r r r@{}}
Name              & Type                & $|V|$ & $|E|$ & $|E|\big/|V|$ & $|S|$ & $|S|\big/|V|$ & $l$   & Source\\\hline
\textsf{lesmis}   & Co-occurrence       & 77    & 254   & 3.30          & 35    & .455          & 2.61  & \small{\cite{konect, StandfordGraphBase}}\\
\textsf{can96}    & Mesh structure      & 96    & 336   & 3.5           &  0    & 0             & 4.36  & \small{\cite{SparseMatrixCollection}}\\
\textsf{football} & Interaction network\hspace{-3pt} & 115   & 613   & 5.33          &  0    & 0             & 2.49  & \small{\cite{konect, Girvan02}}\\
\textsf{rajat11}  & Circuit simulation  & 135   & 276   & 2.04          &  6    & .044          & 5.57  & \small{\cite{SparseMatrixCollection}}\\
\textsf{jazz}     & Collaboration       & 198   & 2,742 & 13.85         & 14    & .071          & 2.22  & \small{\cite{konect, Jazz}}\\
\textsf{netsci}   & Coauthorship        & 379   & 914   & 2.41          & 183   & .483          & 6.03  & \small{\cite{konect, Newman06}}\\
\textsf{dwt419}   & Mesh structure      & 419   & 1,572 & 3.75          & 32    & .076          & 8.97  & \small{\cite{SparseMatrixCollection}}\\
\textsf{asoiaf}   & Co-occurrence       & 796   & 2,823 & 3.55          & 170   & .214          & 3.41  & \small{\cite{konect}}\\
\textsf{bus1138}  & Power system        & 1,138 & 1,458 & 1.28          & 16    & .014          & 12.71 & \small{\cite{SparseMatrixCollection}}
\end{tabu}
}
\label{tab:data}
\vspace{-15pt}
\end{table}
\subsection{Datasets}
\label{sec:eval:data}
We use nine real-world graphs and 20,000 layouts per graph in our evaluations.
For the quantitative reconstruction evaluation, as we perform 5-fold cross-validations, 16,000 layouts are used as the training set, and 4,000 layouts are used as the test set.

\paragraph{\textbf{Graphs}}
\autoref{tab:data} lists the graphs used for our evaluations.
These include varying sizes and types of networks.
We collected the graphs from publicly available repositories \cite{konect, SparseMatrixCollection}.
As disconnected components can be laid out independently, 
we use the largest component if a graph has multiple disconnected components (\textsfsm{netsci}). 
\textsfsm{can96} and \textsfsm{football} do not have any structurally equivalent nodes. 
Therefore, we did not use the Gromov--Wasserstein distance \cite{Memoli11}.

\paragraph{\textbf{Layouts}}
We collected 20,000 layouts for each graph using the four different layout methods and 5,000 different parameter settings per method, where each parameter value is randomly assigned following random search (\autoref{sec:approach:training_data}).
The layout methods and their parameter ranges used in the evaluations are listed in \autoref{tab:layout_methods}.

While there is a plethora of layout methods, 
we selected these four layout methods because they are capable of producing diverse layouts by using the wide range of parameter values.
Also, their publicly available, robust implementations did not produce any degenerate case in our evaluations.
Lastly, these methods are efficient for computing a large number of layouts in a reasonable amount of time.

The resulting 20,000 layouts vary in many ways ranging from aesthetically pleasing looks to incomprehensible ones (e.g., hairballs).
We did not remove any layout as we wanted to observe how the models encapsulate the essence of graph layouts from the diverse layouts.

\subsection{Models and Configurations}
We compare eight different model designs to investigate the effects of graph neural networks (GNNs) and Gromov--Wasserstein (GW) \cite{Memoli11} distance in our model.
All the models we use have the same architecture as described in \autoref{sec:arch}, except for the GNN layers.

\textbf{MLP}: 
This model uses 1-layer perceptrons instead of GNNs; it does not consider the structure of the input graph.
The model serves as a baseline for the investigation of the representational power of GNNs.

\textbf{GCN}: 
This model uses graph convolutional networks (GCN) \cite{Kipf16GCN} as the GNN layers. 
GCN is one of the early works of GNN.

\textbf{GIN-1}: 
This model uses graph isomorphism networks (GIN) \cite{Xu19GIN} as the GNN layers. 
1-layer perceptrons are used as the MLP in \autoref{eq:gin}.

\textbf{GIN-MLP}: 
This model uses GIN \cite{Xu19GIN} as the GNN layers and 2-layer perceptrons as the MLP in \autoref{eq:gin}.
We use the element-wise mean for aggregating the representation of neighbors of a node in both the GIN-1 and GIN-MLP models ($f$ in \autoref{eq:gin}).

Also, for the models that use the GW distance in the optimization process, we add `+GW' to its model name. 
For example, GIN-MLP+GW denotes a model uses GIN-MLP as the GNN layers and the GW distance for comparing the structurally equivalent nodes (SENs) in the optimization process.
The models with GW \cite{Memoli11} are only used for the graphs having SENs (all the graphs except \textsfsm{can96} and \textsfsm{football}).

We use the $L_1$ loss function for the reconstruction loss.
For the variational loss,
we draw the same number of samples as the batch size from the prior distribution and
set $c(x, y) = \|x - y\|_2^2$ to compute the sliced-Wasserstein distances (\autoref{eq:sw}).
Also, we set $\beta = 10$ in the optimization objective (\autoref{eq:loss}).

We use varying numbers of hidden units in GNN layers depending on the number of nodes: 32 units for \textsfsm{lesmis}, \textsfsm{can96}, \textsfsm{football} and \textsfsm{rajat11}, 64 units for \textsfsm{jazz}, \textsfsm{netsci}, and \textsfsm{dwt419}, and 128 units for \textsfsm{asoiaf} and \textsfsm{bus1138}. 
The batch size also varies: 40 layouts per batch for \textsfsm{asoiaf} and \textsfsm{bus1138}, and 100 for the other graphs.

All the models use 
three GNN layers (or perceptrons in the MLP models),
the exponential linear unit (ELU) \cite{ELU} as the non-linearity, 
batch normalization \cite{BatchNorm} on every hidden layers,
and an element-wise mean pooling as the readout function in the encoder.
We use the Adam optimizer \cite{Adam} with a learning rate of 0.001 for all the models.
We train each model for 50 epochs.

\begin{table}[t]
\setlength{\tabcolsep}{0.3em} 
\captionsetup{farskip=0pt,skip=3pt}
\centering
\caption{The layout methods and parameter ranges used for producing training data in our experiments. 
The parameter names follow the documentation of the implementations.}
{\scriptsize
\begin{tabu}{@{}X[4,l] X[22,l] X[2,r]@{}}
Method & Parameter Ranges & \hspace{-10em}Implementation \\\hline
\textsf{D3}\cite{D3} & link~distance~$= [1.0,~100.0]$, charge~strength~$= [-100.0,~-1.0]$, velocity~decay~$= [0.1,~0.7]$ & \cite{d3-force} \\\hline
\textsf{FA2}\cite{ForceAtlas2} & gravity~$= [1.0,~10.0]$, scaling~ratio~$= [1.0,~10.0]$, adjust~sizes~$= \{\text{true},~\text{false}\}$,~linlog~$= \{\text{true},~\text{false}\}$, outbound~attraction~distribution~$= \{\text{true},~\text{false}\}$, strong~gravity~$= \{\text{true},~\text{false}\}$ & \cite{Gephi} \\\hline
\textsf{FM\textsuperscript{3}}\cite{FM3} & force~model~$= \{\text{new},~\text{fr}\}$, galaxy~choice~$=~\{\text{lower~mass},~\text{higher~mass},~\text{uniform}\}$, spring~strength $= [0.1,~1000.0]$, repulsive~spring~strength~$=~[0.1,~1000.0]$, post~spring~strength~$=~[0.01,~10.0]$, post~repulsive~spring~strength~$=~[0.01,~10.0]$ & \cite{OGDF} \\\hline
\textsf{sfdp}\cite{Hu05} & repulsive~force~strength~(\texttt{C})~$= (0.0, 5.0]$, repulsive~force~exponent~(\texttt{p})$~= [0.0, 5.0]$, attractive~force~strength~(\texttt{mu})~$= (0.0, 5.0]$, attractive~force~exponent~(\texttt{mu\_p})~$~= [0.0, 5.0]$ & \cite{graph-tool} \\
\end{tabu}
}
\label{tab:layout_methods}
\vspace{-15pt}
\end{table}

\subsection{Implementation}
We implemented our models in PyTorch \cite{PyTorch}.
The machine we used to generate the training data and to conduct the evaluations has an Intel i7-5960X (8 cores at 3.0 GHz) CPU and an NVIDIA Titan X (Maxwell) GPU.
The implementation of each layout method used in the evaluations is also shown in \autoref{tab:layout_methods}.

\subsection{Test Set Reconstruction Loss}
\label{sec:eval-1}
We compare the test layout reconstruction to evaluate the models' generalization capability.
Here, a layout reconstruction means the model takes the input layout, encodes it to the latent space, and then reconstructs it from the latent representation.
The test reconstruction loss quantifies the generalization ability of a model because it measures the accuracy of reconstructing the layouts that the model did not see in the training.
We perform 5-fold cross-validations to compare the eight model designs in terms of their test set reconstruction loss.
To reduce the effects of the fold assignments, we repeat the experiment 10 times and report the mean losses.

\begin{figure}[H]
\captionsetup{farskip=0pt,skip=0pt}
\centering
\includegraphics[width=\columnwidth, keepaspectratio]{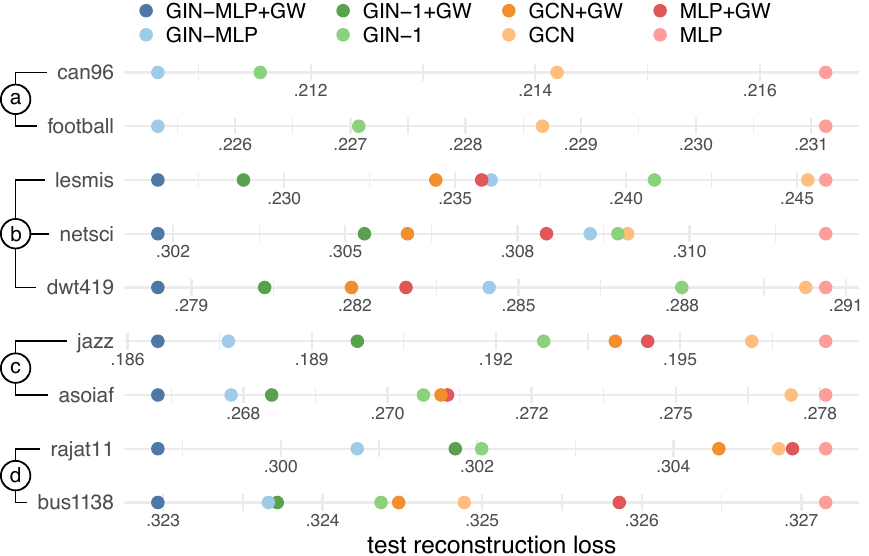}
\vspace{-10pt}
\caption{Average test reconstruction losses.
Since the value ranges vary for each graph, we use individual ranges for each graph.
The models with Gromov--Wasserstein distance \cite{Memoli11} are not used for \textit{can96} and \textit{football} as they do not have any structurally equivalent nodes.}
\label{fig:test_recon_result}
\vspace{-9pt}
\end{figure}

\paragraph{\textbf{Results}}
We compare the eight models in terms of the average reconstruction loss (lower is better) of the test sets, which are the 4,000 layouts that are not used to train the models.
The results shown in \autoref{fig:test_recon_result} are the mean test reconstruction losses of the 10 trials of the 5-fold cross-validations.
The standard deviations are not shown as the values are negligible: all standard deviations are less than $3 \times 10^{-4}$.

Overall, the models with GIN-MLP and GW show the lowest loss of all the graphs.
GIN-MLP models show the lowest loss for the graphs that do not have any structurally equivalent nodes, as shown in \autoref{fig:test_recon_result}a, where the models with GW are not used.
Thus, GIN-MLP+GW models show the lowest loss for all the other graphs (\autoref{fig:test_recon_result}b--d).

Although the absolute differences vary, the ranking between the models based on the neural network modules is consistent in all the graphs.
The ranking of the models with GW is as follows: GIN-MLP+GW, GIN-1+GW, GCN+GW, and MLP+GW.
The ranking of the models without GW is as follows: GIN-MLP, GIN-1, GCN, and MLP.
In addition, the three different rankings of the models are found for the graphs with structural equivalences (\autoref{fig:test_recon_result}b--d).

\paragraph{\textbf{Discussion}}
In \textsfsm{lesmis}, \textsfsm{netsci}, and \textsfsm{dwt419} (\autoref{fig:test_recon_result}b), all the models with GW show a lower loss than the models without GW.
We have found that many nodes in the three graphs have structural equivalences ($|S|\big/|V|$ in \autoref{tab:data}).
This shows that the GW helps the learning process, especially for the graphs that have many structural equivalences.

For \textsfsm{jazz} and \textsfsm{asoiaf} (\autoref{fig:test_recon_result}c), the models with GIN show better results than the models without GIN.
Although they also have a considerable ratio of structural equivalences, 
GIN-MLP and GIN-1 show lower reconstruction losses than GCN+GW and MLP+GW.
Considering the number of nodes as shown in \autoref{tab:data}, we have found that the two graphs (\textsfsm{jazz} and \textsfsm{asoiaf}) are dense and have a relatively small average path length compared to the other graphs.
This suggests the GIN is important for dense and ``small-world''-like networks~\cite{Watts98}. 

For \textsfsm{rajat11} and \textsfsm{bus1138} (\autoref{fig:test_recon_result}d), the ranking of the models is as follows: GIN-MLP+GW, GIN-MLP, GIN-1+GW, GIN-1, GCN+GW, GCN, MLP+GW, and MLP.
We have found that these two graphs are sparse and they have a small ratio of nodes with structural equivalence.
This suggests that the representational power of the GNNs has a stronger effect than the usage of GW in such graphs.
\vspace{-2pt}
\subsection{Layout Metrics and Test Reconstruction Loss}
\label{sec:eval-2}
\vspace{-1pt}
To investigate our models' behavior on reconstructing the test sets,
we analyze the correlations between the test reconstruction loss and each of the two layout quality metrics of the test input layouts: 
crosslessness \cite{Purchase02} and shape-based metric \cite{Eades17}.
Crosslessness~\cite{Purchase02} is a normalized form of the number of edge crossings where a higher value means fewer edge crossings.
Shape-based metric~\cite{Eades17} measures the quality of a layout based on the similarity between the graph and its shape graph (a relative neighborhood graph of the node positions). 
We use the Gabriel graph
\cite{Gabriel69} as the shape graph of a layout.
We use the models with the lowest test reconstruction loss for each graph from \autoref{sec:eval-1}. 

\paragraph{\textbf{Results}}
The Pearson correlation coefficients show 
that the test loss has strong negative correlations with
both the crosslessness ($c$) and the shape-based metric ($s$) 
(i.e., when the reconstruction loss of a test input layout is low then its layout metrics are high).
All the results are statistically significant as all the $p$-values are less than $2.2 \times 10^{-16}$:
\begin{table}[H]
\vspace{-13pt}
\setlength{\tabcolsep}{0.24em}
\captionsetup{farskip=0pt,skip=0pt}
\centering
{\scriptsize
\begin{tabu}{@{}l c c c c c c c c c@{}}
& \textsfverysm{lesmis} & \textsfverysm{can96} & \textsfverysm{football} & \textsfverysm{rajat11} & \textsfverysm{jazz} & \textsfverysm{netsci} & \textsfverysm{dwt419} & \textsfverysm{asoiaf} &\textsfverysm{bus1138}\\\hline
$c$ & $-.69355$ & $-.704$ & $-.703$ & $-.521$ & $-.877$ & $-.488$ & $-.586$ & $-.782$ & $-.517$ \\ 
$s$ & $-.69364$ & $-.812$ & $-.350$ & $-.707$ & $-.749$ & $-.630$ & $-.715$ & $-.311$ & $-.706$ \\ 
\end{tabu}
}
\vspace{-12pt}
\end{table}

\paragraph{\textbf{Discussion}}
The results show that our models learn better on how to reconstruct the layouts with fewer edge crossings and are similar to its shape graph.
In other words, our models tend to generate ``good'' layouts in terms of the two metrics, even if the models are not trained to do so.
We show detailed examples in \autoref{sec:eval-qual}.

\begin{figure*}[t]
\captionsetup{farskip=0pt,skip=0pt}
\centering
\includegraphics[width=\textwidth, keepaspectratio]{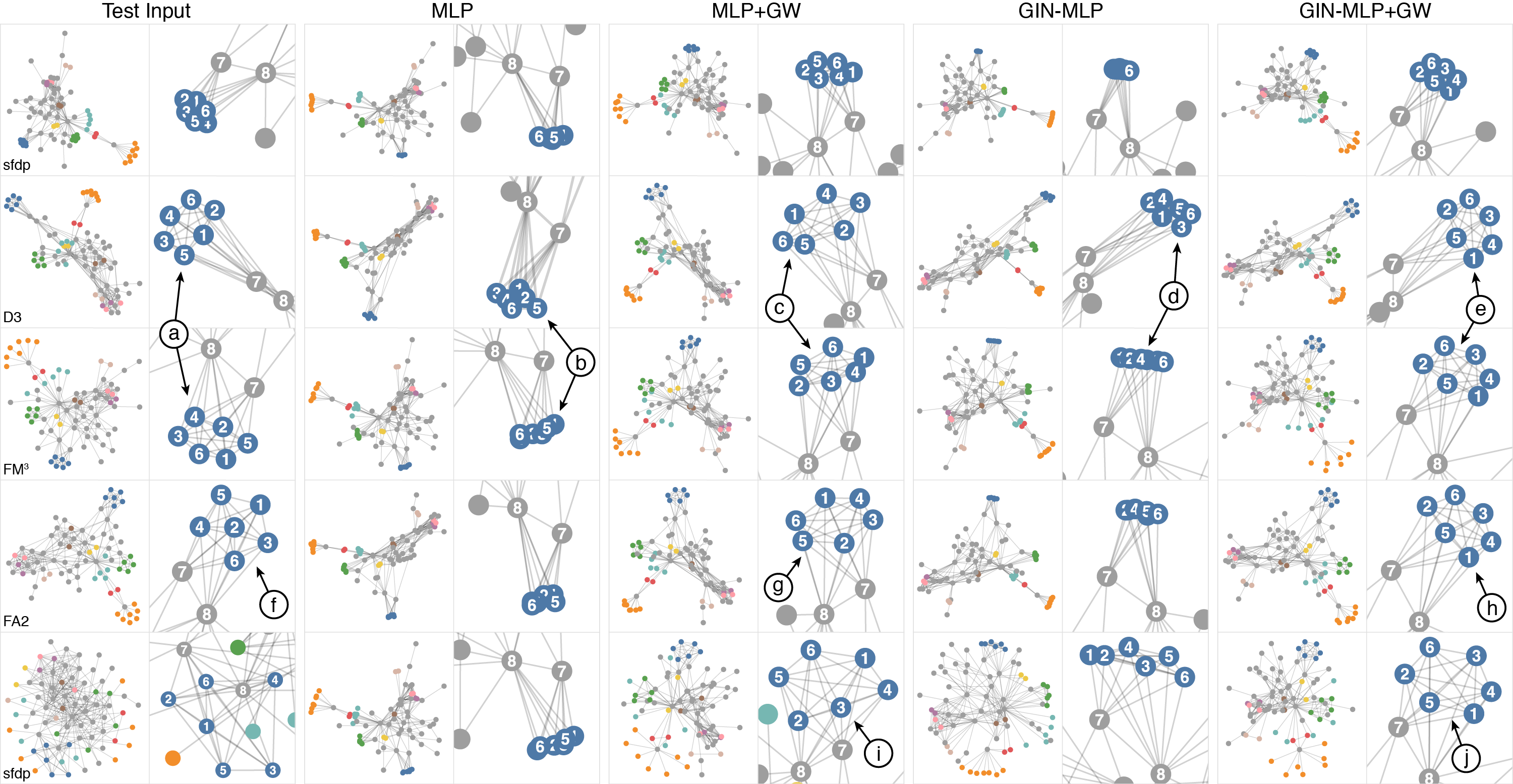}
\vspace{-9pt}
\caption{Qualitative results of the \textit{lesmis} graph using the four different models.
The leftmost column shows the test input layouts that the models did not see in their training session.
The test input layouts are computed with randomly assigned parameter values as described in \autoref{sec:eval:data}.
The other columns show the reconstructed layouts of the test inputs using the four models.
The nodes with the same color, except gray, are structurally equivalent to each other.
The nodes in gray are structurally unique.
The results are discussed in detail in \autoref{sec:eval-qual}.}
\vspace{-15pt}
\label{fig:recon-lesmis}
\end{figure*}

\subsection{Qualitative Results}
\label{sec:eval-qual}
We show the qualitative results of the layout reconstruction and the learned latent space to discuss the behaviors of the models in detail.

\paragraph{\textbf{GIN-MLP and GW}}
The models with GIN-MLP and GW show the lowest test reconstruction loss in \autoref{sec:eval-1}. 
To further investigate, we compare the reconstructed layouts of the \textsfsm{lesmis} graph, which has the widest range of losses among the different models.
Due to space constraints, we discuss the difference between the four different models (MLP, MLP+GW, GIN-MLP, and GIN-MLP+GW) on the \textsfsm{lesmis} graph. 

\autoref{fig:recon-lesmis} shows the qualitative results of reconstructing unseen test input layouts, where a good generative model can produce diverse layouts similar to the unseen input layouts.
The \textsfsm{lesmis} graph has a number of sets of SENs.
In \autoref{fig:recon-lesmis}, the nodes with the same color, except gray, are structurally equivalent to each other; they have the same set of neighbors in the graph. 
The nodes in gray are structurally unique; each of them has a unique relationship to the other nodes in the graph.
For example, the blue nodes (1--6) have the same relationship where they are all connected to each other and to the two other nodes (7 and 8).

As described in \autoref{sec:struct_eq}, the locations of SENs are often not consistent in different layouts.
For example, the arrangements of the blue nodes in the D3 layout and the FM\textsuperscript{3} layout are different (\autoref{fig:recon-lesmis}a).
In the D3 layout, the blue nodes are placed in the following clockwise order: 2, 1, 5, 3, 4, and 6, where the nodes $\{2, 1, 5\}$ are closer to $\{7, 8\}$ than $\{3, 4, 6\}$.
However, in the FM\textsuperscript{3} layout, the clockwise order is 4, 2, 5, 1, 6, and 3, where the nodes $\{4, 2, 5\}$ are closer to $\{7, 8\}$ than $\{1, 6, 3\}$.

The arrangements of the SENs in the reconstructed layouts vary depending on the model.
For example, the models with GW are able to lay out the blue nodes (\autoref{fig:recon-lesmis}c and e) similar to the input layouts (\autoref{fig:recon-lesmis}a).
However, the models without GW fail to learn this and produce collapsed placements (\autoref{fig:recon-lesmis}b and d).
We suspect this is due to the many possible permutations between a set of SENs. The models without GW tend to place a set of SENs at the average position of the SENs to reduce the average loss.
However, the models with GW learn about the generalized concept of the blue nodes' arrangements and produce similar arrangements as the test input layouts in a different permutation of the SENs.
The other sets of SENs show similar results (e.g., the orange and green nodes in \autoref{fig:recon-lesmis}), where the models without GW produce collapsed arrangements, but the models with GW produce similar arrangements as the test input layouts.

In addition, the placements of the blue nodes are spatially consistent across the different layouts using GIN-MLP+GW (\autoref{fig:recon-lesmis}e, h, and j) than using MLP+GW (\autoref{fig:recon-lesmis}c, g, and i).
This shows that the models with GIN-MLP gain a more stable generalization of the arrangement of SENs than the models with only MLP.
Therefore, using GIN-MLP+GW models, users can generate diverse layouts while preserving their mental map across different layouts.
This is not possible in many existing layout methods due to their nondeterministic results \cite{Munzner14:VAD}. 

There are other examples of the generalization capability of our models.
For example, \autoref{fig:recon-lesmis}f shows a different arrangement of the blue nodes from \autoref{fig:recon-lesmis}a, where the layout in \autoref{fig:recon-lesmis}f has a star-like arrangement with node 2 in the center.
However, the reconstructed layouts (\autoref{fig:recon-lesmis}h) of the layout in \autoref{fig:recon-lesmis}f are more similar to the layouts in \autoref{fig:recon-lesmis}a.
We have found that arrangements similar to the layouts in \autoref{fig:recon-lesmis}a are more dominant in the training set.

Another example is the last row of the \autoref{fig:recon-lesmis}, where the input layouts are hairballs.
In contrast, the reconstructed layouts using our models are more organized.
This also shows that the models have a generalized concept of graph layouts.
This might explain the negative correlations in \autoref{sec:eval-2} as the models do not reconstruct the same hairball layouts.

\begin{figure*}[htb]
\captionsetup{farskip=0pt,skip=0pt}
\centering
\includegraphics[width=\textwidth, keepaspectratio]{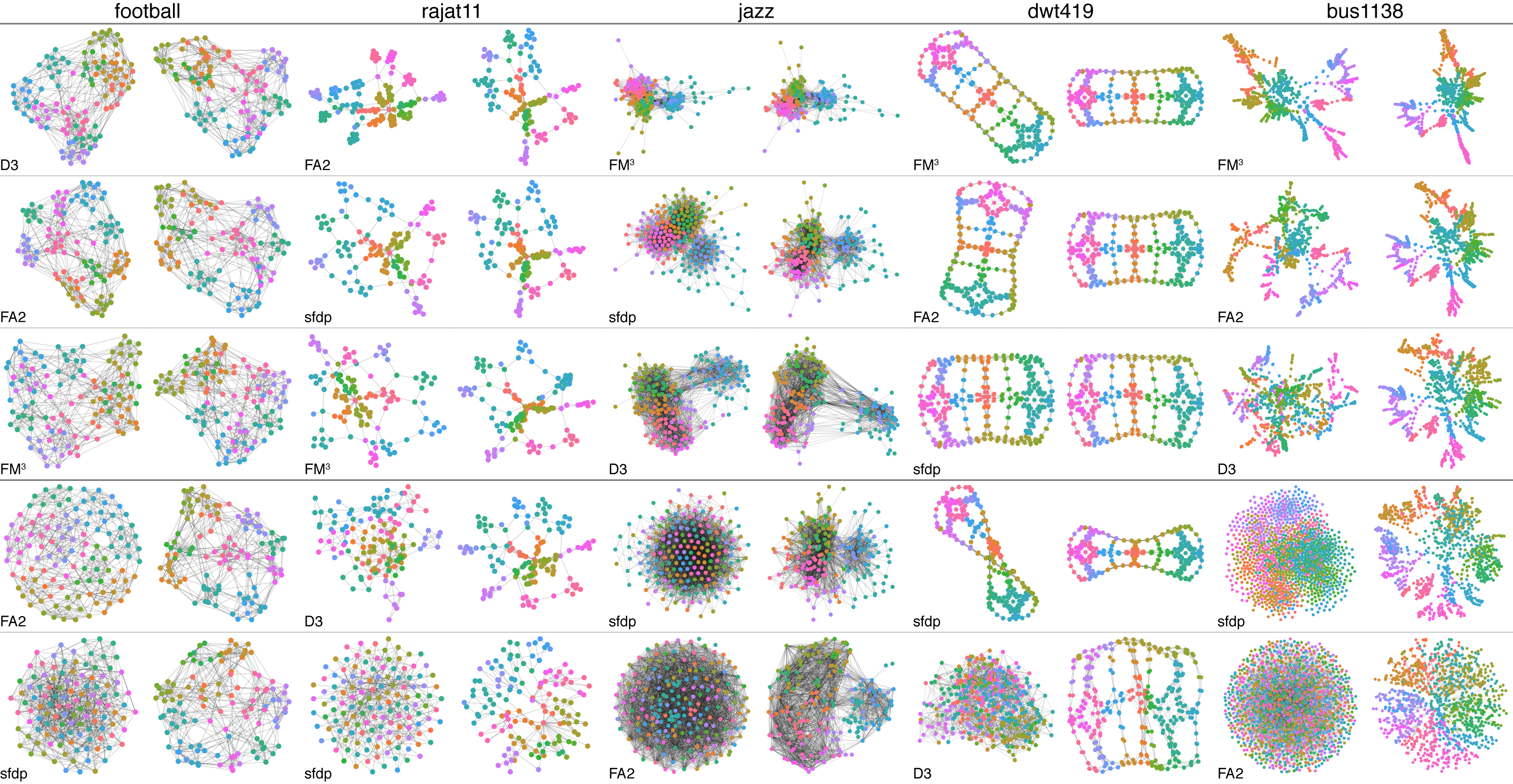}
\vspace{-10pt}
\caption{%
Qualitative results of the five different graphs.
The GIN-MLP model is used for the \textit{football} graph, and the GIN-MLP+GW models are used for the other graphs.
For each pair of layouts, the left is the test input, and the right is the reconstructed layout.
The first three rows show the different styles of layouts for each graph, and the bottom two rows show the reconstruction results of hairball layouts.
The results are discussed in \autoref{sec:eval-qual}.
}
\vspace{-15pt}
\label{fig:recon-gin-mlp+gw}
\end{figure*}

\paragraph{\textbf{Multiple Graphs}}
To demonstrate our models with more graphs, 
\autoref{fig:recon-gin-mlp+gw} shows the qualitative results of five different graphs using the GIN-MLP model for the \textsfsm{football} graph and the GIN-MLP+GW models for the other graphs.
The first three rows show that our models are capable of learning different styles of layouts.
The bottom two rows show the models' behavior on hairball-like input layouts.
We can see that the reconstructed layouts are more organized.
For example, the fourth layout of \textsfsm{dwt419} is twisted. 
However, the reconstructed layout is not twisted but pinched near the center, like other layouts.
These examples might explain the negative correlations in \autoref{sec:eval-2} as the models do not reconstruct hairball layouts well in favor of generalization.

\paragraph{\textbf{Latent Space}}
A common way to qualitatively evaluate a generative model is to show the interpolations between the different latent variables \cite{GAN, VAE, SWAE, MusicVAE, SketchRNN}.
If the transitions between the generated samples based on the interpolation in the latent space are smooth, we can conclude that the generative model has a generalization capability of producing new samples that were not seen in training.

\begin{figure*}[htb]
\captionsetup{farskip=0pt,skip=0pt}
\centering
\includegraphics[width=\textwidth, keepaspectratio]{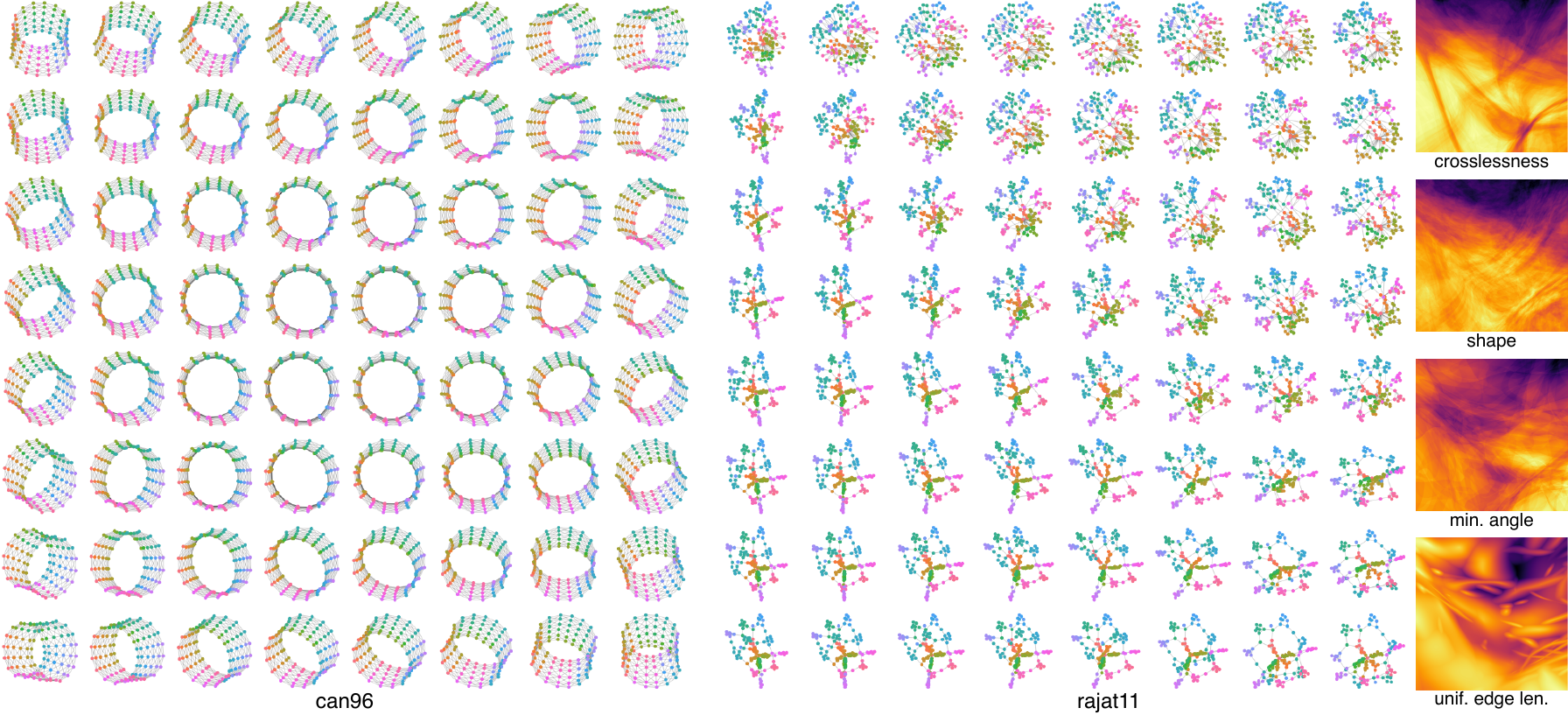}
\vspace{-8pt}
\caption{%
Visualization of the latent spaces of \textit{can96} and \textit{rajat11}.
The GIN-MLP model is used for \textit{can96} and the GIN-MLP+GW model is used for \textit{rajat11}.
The grids of layouts are the generated samples by the decoding of a 8 $\times$ 8 grid in $[-1, 1]^2$.
Also, \autoref{fig:teaser} shows the sample grid of \textit{lesmis}.
The smooth transitions between the generated layouts show that the capability of generalization of our models. 
Novices can directly use this as a WYSIWYG interface to generate a layout they want.
The rightmost column shows the heatmaps of the four layout metrics of 540 $\times$ 540 generated sample layouts in $[-1, 1]^2$ of \textit{rajat11}. 
Experts can use the heatmaps to see complex patterns of layout metrics on diverse layouts.
The results are discussed in detail in \autoref{sec:eval-qual} and \autoref{sec:scenario}.
The results of other graphs and models are available in the supplementary material \cite{Supplementary}.
}
\vspace{-12pt}
\label{fig:latent-space}
\end{figure*}

As our models are trained to construct a two-dimensional latent space, we can show a grid of generated samples interpolating throughout the latent space.
We show the results of \textsfsm{can96} and \textsfsm{rajat11} in \autoref{fig:latent-space} and the result of the \textsfsm{lesmis} in \autoref{fig:teaser}.
As we can see, the transitions between the generated samples are smooth. 
Also, the latent space of \textsfsm{can96} is particularly interesting. 
It seems that the model learned to generate different rotations of the 3D mesh. 
However, all of the layouts used in this paper are 2D. 
In addition, the input feature of a layout is the normalized pairwise distances between the nodes, as described in \autoref{sec:layout-feature}, which do not explicitly convey any notion of 3D rotation.

Based on these findings, we conclude that our models are capable of learning the abstract concepts of graph layouts and generating diverse layouts.
An interactive demo is available in the supplementary material \cite{Supplementary} for readers to explore the latent spaces of all the graphs using all the models we have described in this paper.
\begin{figure}[b!]
\vspace{-12pt}
\captionsetup{farskip=0pt,skip=0pt}
\centering
\subfloat[]{\setlength{\tabcolsep}{0.3em}
{\scriptsize
\begin{tabular}{@{}l r r r r r@{}}
Name              & D3   & FA2  & FM$^3$ & sfdp & Epoch \\\hline
\textsf{lesmis}   & .195 & .620 & .015   & .229 & 76.7 \\
\textsf{can96}    & .239 & .695 & .021   & .329 & 34.7 \\
\textsf{football} & .302 & .845 & .033   & .486 & 43.9 \\
\textsf{rajat11}  & .358 & .856 & .044   & .586 & 45.1 \\
\textsf{jazz}     & .599 & 1.61 & .116   & 1.18 & 212  \\
\textsf{netsci}   & 1.14 & 2.50 & .236   & 2.59 & 281  \\
\textsf{dwt419}   & 1.24 & 3.08 & .257   & 3.08 & 211  \\
\textsf{asoiaf}   & 3.03 & 9.42 & .620   & 9.16 & 690  \\
\textsf{bus1138}  & 4.41 & 7.88 & 1.07   & 10.3 & 509 
\end{tabular}}}\hfill
\subfloat[]{\raisebox{-.5\height}{\includegraphics[]{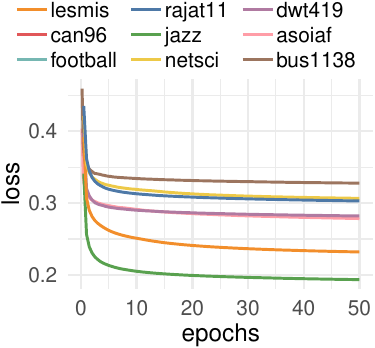}}}
\vspace{-12pt}
\caption{%
Computation time.
The left table shows the layout computation times (D3, FA2, FM\textsuperscript{3}, and sfdp) for collecting the training data and training the model (Epoch).
The layout computation times are the mean seconds for computing one layout per method per graph.
The training computation times are the mean seconds for training one epoch (16K samples) per graph.
The right chart shows that the average training loss is updated every batch to demonstrate that our models converge quickly.}
\vspace{-1.5pt}
\label{fig:comp-time}
\end{figure}

\subsection{Computation Time}
We report and discuss the layout computation time, model training time, and layout generation time of each graph using the models with the lowest test reconstruction loss in \autoref{sec:eval-1}.
The layout computation time and model training time are shown in \autoref{fig:comp-time}.

As we collect a large number of layouts (16K samples per graph for training), computing layouts is the most time-consuming step in building the models.
However, we can incrementally generate training examples and train the model simultaneously.

We have found that the number of nodes having structural equivalence ($|S|$ in \autoref{tab:data}) is a strong factor to the training time.
It is directly related to the complexity of the GW distance \cite{Memoli11} computation (\autoref{sec:struct_eq}).
If we can build a generative model using GANs \cite{GAN}, instead VAEs \cite{VAE} (SWAE \cite{SWAE} in this paper), we can remove the GW distance computation in the process (more in \autoref{sec:discussion}). 
This can significantly reduce the training time.
Also, the number of edges might be a stronger factor to the training time than the number of nodes as we have implemented GNNs using sparse matrices.
For a small graph (e.g., $|V| < 200$), it might be faster to use dense matrices for GNNs.

In addition, the models converge quickly as shown in \autoref{fig:comp-time}. 
Although the models are trained for 50 epochs for our evaluations, 
the models are capable of generating diverse layouts less than 10 epochs.

After training, the mean layout generation times for all the graphs are less than .003 s. 
Thus, users can explore and generate diverse layouts in real time, which is demonstrated in the supplementary material \cite{Supplementary}.

Based on these findings, we expect training a new model for a graph from scratch without any layout examples can be done within a few minutes for a small graph ($|V| < 200,\ |E| < 500$) and a few hours for larger graphs.
Although it is a considerable computation time, our model can be trained in a fully unsupervised manner; it does not require users to be present during the training.
Thus, our approach can save the user's time---which is much more valuable than the computer's time---by preventing them from blindly searching for a good layout.

\section{Usage Scenario}
\label{sec:scenario}
The previous section shows that our model is capable of generating diverse layouts by learning from existing layouts.
This section describes how users can use the trained model to produce a layout that they want.

After training, users can generate diverse layouts of the input graph by feeding different values of the latent variable ($z_L$ in \autoref{sec:arch}) to the decoder.
Thus, the interpretability of the latent space is important for users to easily produce a layout that they want.
We achieve this by using a 2D latent space rather than a higher-dimensional space.
A 2D latent space is straightforward to map additional information onto it.

By mapping generated samples on the 2D latent space (e.g., the sample grid in \autoref{fig:teaser} and \autoref{fig:latent-space}), we can build a what-you-see-is-what-you-get (WYSIWYG) interface for users to intuitively produce a layout that they want.
Finding a desired layout from an unorganized list of multiple layouts (e.g., the training samples in \autoref{fig:teaser}) often results in a haphazard and tedious exploration \cite{Biedl98}.
However, the sample grid provides an organized overview with a number of representative layouts of the input graph.
With the sample grid as a guide, users can intuitively set the latent variable ($z_L$) to produce a suitable layout by pointing a location in the 2D latent space.
They also can directly select the desired one from the samples.
The WYSIWYG interfaces of each graph are demonstrated in the supplementary material \cite{Supplementary}.

Moreover, our approach produces spatially stable layouts.
As we have discussed in \autoref{sec:layout-feature}, many layout methods are nondeterministic. 
For example, in the training samples of \autoref{fig:teaser}, the locations of the orange nodes vary greatly across different layouts.
Thus, identifying the same node(s) among these layouts is difficult because region-based identifications cannot be utilized \cite{Munzner14:VAD}.
Comparing nondeterministic layouts often requires a considerable amount of the user's time since they need to match the nodes between different layouts.
However, as shown in \autoref{fig:teaser} and \autoref{fig:latent-space}, our models produce spatially stable layouts, where the same node is placed in similar locations across different layouts.
Hence, identifying the same node(s) in different layouts is straightforward, and thus comparing layouts becomes an easy task.

Using our approach, users can directly see what the layout results will look like with the sample grid. 
Also, spatially-stable layout generation enables users to effortlessly compare various layouts of the input graph.
Thus, users can intuitively produce a layout that best suits their requirements (e.g., highlighting the interconnections between different communities) 
without blindly tweaking parameters of layout methods.

By mapping layout metrics on the latent space,
users can directly see the complex patterns of the metrics on diverse layouts of a graph.
\autoref{fig:latent-space} (the rightmost column) shows heatmaps of four layout metrics of 540 $\times$ 540 layouts of \textsfsm{rajat11}.
While the sample grid shows smooth transitions between different layouts, the heatmaps show interesting patterns.
For example, there are several steep ``valleys'' in the heatmap of crosslessness \cite{Purchase02}, where the darker colors mean more edge crossings.
This shows crosslessness is sensitive to certain changes in the layouts.
Using the heatmap as an interface, users can exactly see these changes through producing several layouts by pointing the locations across the valleys in the heatmap.
Thus, experts in graph visualization can use the heatmaps for designing layout metrics as they can understand how layout metrics behave on various layouts with concrete examples.

Our approach is an example of \textit{artificial intelligence augmentation} \cite{AIA}, where our generative model builds a new type of user interface with the latent space to augment human intelligence tasks, such as creating a good layout and analyzing the patterns of layout metrics.

\section{Discussion}
\label{sec:discussion}
\autoref{sec:eval} and \autoref{sec:scenario} have discussed the evaluation results and usage scenarios.
This section discusses the limitations and future research directions of our approach.
This paper has introduced the first approach to generative modeling for graph layout.
As the first approach in a new area, there are several limitations we hope to solve in the future.

The maximum size of a graph in our approach is currently limited by the capacity of GPU memory.
This is the reason we could not use a larger batch size for \textsfsm{asoiaf} and \textsfsm{bus1138} (40 layouts per batch, while 100 layouts per batch for the other graphs).
As a sampling-based GNN \cite{SSE} scalable to millions of nodes has been recently introduced, 
we expect our approach can be applied to larger graphs in the future.

Although our model is capable of generalizing for different layouts of the same graph, it does not generalize for both different graphs and different layouts.
Therefore, we need to train a new model for each graph.
Our model can be trained in a fully unsupervised manner and can be trained incrementally while generating the training samples simultaneously.
A better model would learn to generalize across different graphs so that it can be used for any unseen graphs.
However, this is a very challenging goal.
Most machine learning tasks that require the generalization across different graphs (e.g., graph classifications) aims to learn graph-level representations.
But the generalization across both graphs and their layouts in a single model requires to learn the latent representations of nodes across many different graphs.
Unfortunately, this is still an open problem.

Our model learns the data distribution of the training set.
However, a valid layout can be ignored in favor of generalization.
For example, the arrangement of the blue vertices in \autoref{fig:recon-lesmis}f is a valid layout, 
but it is a rare type of arrangement in the training dataset. 
Thus, the model produces a more general arrangement following the distribution of the training set (\autoref{fig:recon-lesmis}h).
As a valid layout can be an outlier in the training dataset,
we need an additional measure to not over-generalize valid layouts in the training dataset and properly reconstruct valid layouts.

We have used a sliced-Wasserstein autoencoder \cite{SWAE}, a variant of VAE \cite{VAE}, for designing our architecture. 
As a VAE, we explicitly define a reconstruction loss for the training.
However, this was challenging for comparing two different layouts of a set of structurally equivalent nodes.
In this paper, we have used the GW distance \cite{Memoli11} to address this issue.
Another possible solution is to use GANs, which do not require an explicit reconstruction loss function in the optimization process.
Thus, using a GAN can reduce the computational cost because computing GW distance is no longer required.
While we did not use GANs due to mode collapse and non-convergence \cite{GANTutorial}, we believe it is possible to use GANs for graph layout in the future.

In this work, we map a number of generated samples on a 2D latent space to directly see the latent space.
However, the learned latent representation of our model is \textit{entangled}, which means each dimension of the latent space is not interpretable.
As we can see from the grids of samples in \autoref{fig:teaser} and \autoref{fig:latent-space}, although we can see what the generated layouts look like with different latent variables, we cannot interpret the meaning of each dimension.
Thus, it is difficult to use a higher-dimensional latent space for our purpose, as we cannot either interpret each dimension or see the overview of the latent space.
Learning a model that produces disentangled representations is an important research direction in generative modeling \cite{BVAE}.
With a generative model that can learn a disentangled latent space, we can produce a layout in a more interpretable way, where each dimension only changes a specific aspect of a layout independently.
For example, if one dimension of the latent representation encodes the area of clusters of nodes, we can directly manipulate a layout to change the cluster size of a layout.

\section{Conclusion}
Graph-structured data is one of the primary classes of information. 
Creating a good layout of a graph for visualization is non-trivial.
The large number of available layout methods and each method's associated parameter space confuse even the experts.
The trial-and-error efforts require a significant amount of the user's time.
We have introduced a fundamentally new approach to graph visualization, where we train a generative model that learns how to visualize a graph from a collection of examples.
Users can use the trained model as a WYSIWYG interface to effortlessly generate a desired layout of the input graph.

Generative modeling for image datasets has shown dramatic performance improvement;
it took only four years from the first model of generative adversarial networks \cite{GAN} to an advanced model that can generate high-resolution images \cite{ProgressiveGAN}.
There can be many exciting ways to use generative models 
for graph visualization, or even other types of data visualization.
We hope this paper will encourage others to join this exciting area of study to accelerate designing generative models for revolutionizing visualization technology.

\acknowledgments{This research has been sponsored in part by the U.S. National Science Foundation through grant IIS-1741536.}

\bibliographystyle{abbrv}

\bibliography{tex/references}
\end{document}